\documentclass[a4paper,aps,twocolumn]{revtex4}

\begin{document}

\title{Phase space flows for 
non-Hamiltonian systems with constraints}

\author{Alessandro Sergi
\footnote{ (E-mail: asergi@unime.it)}}
\affiliation{
Dipartimento di Fisica, Sezione Fisica Teorica,
Universit\'a degli Studi di Messina, 
Contrada Papardo C.P. 50-98166 Messina, Italy
}

\begin{abstract}
In this paper, non-Hamiltonian systems with holonomic constraints 
are treated by a generalization of Dirac's formalism.  
Non-Hamiltonian phase space flows can be described
by generalized antisymmetric brackets
or by general Liouville operators which cannot be derived
from brackets.  Both situations are treated.
In the first case, a Nos\'e-Dirac 
bracket is introduced as an example.  
In the second one, Dirac's recipe for
projecting out constrained variables from time translation operators
is generalized and then applied to non-Hamiltonian linear response.
Dirac's formalism avoids spurious terms in the response function
of constrained systems. However, corrections coming from phase
space measure must be considered for general perturbations.
\end{abstract}

\maketitle


\section{Introduction}

Constrained systems are ubiquitous in theory and computation
and formalisms for their treatment are
still developed~\cite{minary,minary2,sergi-dirac}.
Some time ago Dirac showed how to formulate generalized
Hamiltonian phase space flows which automatically
satisfy certain class of 
constraints~\cite{dirac1,dirac2,sudarshan,regge,sundermeyer}.
These constraints, which are called {\sl second class},
are specified by a non-zero Poisson bracket with
the Hamiltonian. Dirac's investigation aimed
at finding a quantization procedure for relativistic
fields which have constraints arising from their
Lorentz or gauge symmetries.
Recently it has been shown~\cite{sergi-dirac}
how Dirac's scheme can be also applied 
to non-relativistic systems,  such as those
addressed by classical molecular dynamics simulations 
in condensed matter, where constraints are used
to describe the topology of molecules or rare events~\cite{bm}.
In particular it has been shown how Dirac's formalism
can be subsumed by means of a generalized bracket
introduced in Refs.~\cite{b1,b2}.

The concern of this paper is to generalize
the approach of~\cite{sergi-dirac} to cases
where the dynamics is 
non-Hamiltonian~\cite{b1,b2,tuck1,tuck2,tuck3,ramshaw,ezra,ezra2}.
This is useful since computational schemes
adopt constraints and non-Hamiltonian dynamics at the same time,
with the latter implementing both specific thermodynamic
conditions, i.e. constant temperature~\cite{nose,nh,nhc,ferrario},
and nonequilibrium perturbations not derivable from a
Hamiltonian~\cite{ladd,sllod,evansmorriss,neqtuck}.
Building on the results given in Ref.~\cite{sergi-dirac},
the discussion will be limited to systems with holonomic constraints.
Nonholonomic constraints (such as those involved in the formulation of the isokinetic
ensemble~\cite{minary,minary2}) will not be addressed in this paper.  
Two kinds of non-Hamiltonian dynamics
will be considered. The first is based on the generalized  brackets
introduced in~\cite{b1,b2} while the second case arises from
Liouville operators which cannot be expressed by means
of antisymmetric brackets.
For the first case, the  antisymmetric structure
of the generalized bracket will be used
to combine Dirac's theory
with non-Hamiltonian Nos\'e-Hoover dynamics
(more general dynamics, such as Nos\'e-Hoover chains~\cite{nhc},
do not introduce any conceptual difference).
As an application of the Nos\'e-Dirac phase
space flow, the unbiasing factor, arising when holonomic
constraints are used to study rare events~\cite{bm},
is re-derived. 
The second type of non-Hamiltonian dynamics
requires a generalization of Dirac's scheme
in order to project out the constrained degrees
of freedom from any arbitraty Liouville operator.
Linear response theory will be reviewed 
and some fine points, which are relevant for analyzing dynamics with holonomic 
constraints (such as in the case 
of molecular systems~\cite{allen,edberg,marechal,travis}), will be discussed.
In particular, it will be shown that correction terms,
stemming from phase space measure, appear
in the response function for general forms 
of perturbations.

The paper is organized as follows. In Sec.~\ref{sec:1},
Dirac's Hamiltonian formalism is briefly reviewed.
In Sec.~\ref{sec:2} a unified bracket for Nos\'e
thermostated dynamics and constraints, producing
a non-Hamiltonian Nos\'e-Dirac phase space flow, is introduced.
As an illustration of the formalism, the Nos\'e-Dirac
flow is applied in Appendix~\ref{sec:app}
to the discussion of rare events sampling.
In Sec.~\ref{sec:3}, Dirac's recipe, for projecting out
the spurious dynamics of constrained variables,
is first generalized 
to arbitrary time-translation operators
and then applied to non-Hamiltonian Liouville
operators which cannot be derived from brackets.
Linear response theory is briefly reviewed,
discussing how Dirac's prescription avoids
fake terms coming from constraints.
Nevertheless, it is shown that corrections terms 
in the response function,
originating from the constrained phase space measure,
may appear in the general case.

\section{Hamiltonian formalism for systems with holonomic constraints}
\label{sec:1}

Consider a system with a conserved energy ${\cal H}_0({\bf x})$,
where ${\bf x}=({\bf r},{\bf p})$  denotes
the phase space point. To formulate phase space equations of
motion in the presence of mechanical constraints one can follow
Dirac's 
approach~\cite{sergi-dirac,dirac1,dirac2,sudarshan,regge,sundermeyer}.
Together with the $n$ constraints in configuration space
$\mbox{\boldmath$\sigma$}({\bf r})=0$, one has to consider an additional number
$n$ of phase space constraints $\dot{\mbox{\boldmath$\sigma$}}({\bf
r},\dot{\bf r})=0$. It is useful to let
$\mbox{\boldmath$\chi$}=(\mbox{\boldmath$\sigma$},\dot{\mbox{\boldmath$\sigma$}})$
denote the entire set of $2n$ phase space constraints. The $n$
constraints $\dot{\mbox{\boldmath$\sigma$}}=0$ are redundant but
necessary to set up a phase space picture of the dynamics.
Following the convention (due to Dirac~\cite{dirac1,dirac2})
of evaluating derivatives first and 
imposing constraint relations after, these constraints will disappear
from the equations of motion and will not contribute to the phase
space measure.

The equations of motion with constraints may be written as~\cite{sergi-dirac}
\begin{equation}
\dot{x}_i=\sum_{j=1}^{2N}{\cal B}_{i j}^D({\bf x})
\frac{\partial{\cal H}_0}{\partial x_j} \;, \label{eq:b1}
\end{equation}
where $2N$ is phase space dimension and
$\mbox{\boldmath ${\cal B}$}^D$ is an antisymmetric tensor
defined by
\begin{equation}
{\cal B}^D_{i j}(x)={\cal B}_{i j}^c-\sum_{k,l}^{2N}\sum_{\alpha,\beta}^{2n}
{\cal B}_{i k}^c\frac{\partial\chi_{\alpha}}{\partial x_k}
\left({\bf C}^{-1}\right)_{\alpha\beta}
\frac{\partial\chi_{\beta}}{\partial x_l}{\cal B}_{l j}^c
\;,
\label{B^D}
\end{equation}
with
\begin{equation}
\mbox{ \boldmath$\cal B$ }^c =\left[\begin{array}{cc}
{\bf 0}&{\bf 1}\\
-{\bf 1}& {\bf 0}\end{array}\right]
\end{equation}
usually called the symplectic matrix~\cite{goldstein}.
In order to arrive at Eq.~(\ref{B^D}) one has to define
\begin{equation}
 C_{\alpha\beta}=\{\chi_{\alpha},\chi_{\beta}\}=\sum_{ij}
\frac{\partial\chi_{\alpha}}{\partial x_i}{\cal B}_{ij}^c
\frac{\partial\chi_{\beta}}{\partial x_j}
\label{eq:c}
\end{equation}
given in
terms of Poisson bracket of the $2n$ phase space constraints,
and the
inverse matrix $\left({\bf C}^{-1}\right)_{\alpha\beta}$
($\alpha,\beta=1,...,2n$),written explicitly in block form
as
\begin{equation}
{\bf C}^{-1}=\left[\begin{array}{cc} \mbox{\boldmath
$Z$}^{-1}\mbox{\boldmath$\Gamma$}~\mbox{\boldmath$ Z$}^{-1} &
-\mbox{\boldmath$Z$}^{-1} \\
\mbox{\boldmath$Z$}^{-1} & {\bf 0} \end{array}\right]\;,
\label{eq:c-1}
\end{equation}
where the matrices 
\begin{equation}
Z_{\alpha\beta}=\sum_{i=1}^N \frac{1}{m_i}\nabla_i\sigma_{\alpha}
\nabla_i\sigma_{\beta}
\end{equation}
and
\begin{eqnarray}
\Gamma_{\alpha\beta}&=&\sum_{i,k=1}^N\frac{p_i}{m_im_k}\Big(
\nabla_{ki}^2\sigma_{\alpha}\nabla_k\sigma_{\beta}-\nabla_k\sigma_{\alpha}
\nabla_{ki}^2\sigma_{\beta}\Big)
\end{eqnarray}
(with $\alpha,\beta=1,\ldots,n$)
have been defined.

The matrix $\mbox{\boldmath ${\cal B}$}^D$ can be written
explicitly in block form as
\begin{equation}
\mbox{\boldmath ${\cal B}$}^D=
\left[\begin{array}{cc}
{\bf 0} &{\bf 1}-\mbox{\boldmath$\Delta$}\\
-{\bf 1}+\mbox{\boldmath$\Delta$}^T & \mbox{\boldmath$\Lambda$}
\label{eq:B^D}
\end{array}\right]\;,
\end{equation}
where 
\begin{eqnarray}
\Delta_{ij}&=&\sum_{k,l=1}^N
\sum_{\alpha,\beta=1}^n
{\cal B}_{i,N+k}^c
\frac{\partial\dot{\sigma}_{\alpha}}{\partial{\bf p}_k}
(\mbox{\boldmath$Z$}^{-1})_{\alpha\beta}
\frac{\partial\sigma_{\beta}}{\partial{\bf r}_l}
{\cal B}_{l,N+j}^c
\nonumber \\
&&
 (i,j=1,\ldots,N)
\end{eqnarray}
and
\begin{eqnarray}
\Lambda_{ij}&=&-\sum_{k,l=1}^N\sum_{\alpha,\beta=1}^n
{\cal B}_{N+i,k}^c
\frac{\partial\sigma_{\alpha}}{\partial{\bf r}_k}
(\mbox{\boldmath$Z$}^{-1}
\mbox{\boldmath$\Gamma$}
\mbox{\boldmath$Z$}^{-1})_{\alpha\beta}
\frac{\partial\sigma_{\beta}}{\partial{\bf r}_l}{\cal B}_{l,N+j}^c
\nonumber \\
&& (i,j=1,\ldots,N) \;.
\end{eqnarray}

Substituting $\mbox{\boldmath ${\cal B}$}^D$ into
Eq.~(\ref{eq:b1}) and taking into account the fact that
$\dot{\mbox{\boldmath$\sigma$}}=0$, one obtains the equations of
motion~\cite{sergi-dirac}
\begin{eqnarray}
\dot{\bf r}_i&=&\frac{{\bf p}_i}{m_i}\label{eq:dotr} \\
\dot{\bf p}_i&=&{\bf F}_i+\nabla_i\mbox{\boldmath$\sigma$}
\cdot\mbox{\boldmath$\lambda$}({\bf r},{\bf p})\;,\label{eq:dotp}
\end{eqnarray}
where
\begin{equation}
\mbox{\boldmath$\lambda$}({\bf r},{\bf p})
=\mbox{\boldmath$Z$}^{-1}\cdot
\{\dot{\mbox{\boldmath$\sigma$}},{\cal H}_0\}\;.
\end{equation}
Equations~(\ref{eq:dotr})-(\ref{eq:dotp}) have a phase space
compressibility~\cite{sergi-dirac,melchionna}
\begin{equation}
\kappa_c=-\frac{d}{dt}\ln||\mbox{\boldmath$Z$}||
\label{eq:k_c}
\end{equation}
and distribution function~\cite{sergi-dirac,tuck3}
\begin{equation}
\rho_e=\delta({\cal H}_0)\delta(\mbox{\boldmath$\chi$})||\mbox{\boldmath$Z$}||
\label{eq:rhoeham}
\end{equation}
where $\delta(\mbox{\boldmath$\chi$})=\prod_{\alpha}
\delta(\mbox{\boldmath$\sigma$}_{\alpha})
\delta(\dot{\mbox{\boldmath$\sigma$}_{\alpha}})$.
It is worth to remark that Eqs.~(\ref{eq:b1}) with the tensor in Eq.~(\ref{B^D}),
and their explicit form (\ref{eq:dotr})-(\ref{eq:dotp}),
can be regarded as Hamiltonian since the associated generalized bracket
\begin{equation}
(a,b)_D=\sum_{i,j=1}^N\frac{\partial a}{\partial x_i}{\cal B}_{ij}^D
\frac{\partial b}{\partial x_j}\;,\label{eq:genbra}
\end{equation}
where $a$ and $b$ are arbitrary phase space functions, satisfies
the Jacobi relation~\cite{sergi-dirac,morrison}.
The generalized bracket in Eq.~(\ref{eq:genbra})
has the property of leaving invariant, by construction,
any function of the constraints.

\section{Nos\'e-Dirac phase space flow}\label{sec:2}

Starting from the structure of either~(\ref{eq:b1}) or the
associated generalized bracket in~(\ref{eq:genbra}),
with {\boldmath$\cal B$}$^D$ given by Eq.~(\ref{B^D}),
it is very easy to define non-Hamiltonian equations
of motion. It suffices to substitute
the tensor {\boldmath$\cal B$}$^c$ in Eq.~(\ref{B^D})
with a more general antisymmetric tensor {\boldmath$\cal B$}$(x)$
so that the Jacobi identity is no longer satisfied.
When one tries to apply this program to extended system dynamics, as in the case 
of Nos\'e-Hoover dynamics, problems are encountered
since the constraints are usually defined only onto a subspace
of the extended phase space variables. This straightforward
approach would make the generalized bracket identically zero.
One can bypass this problem by exploiting the block structure
of {\boldmath$\cal B$}$^D$ as given in~(\ref{eq:B^D}).
To this end, consider Nos\'e extended phase space with coordinates
${\bf x}=({\bf r},{\eta},{\bf p},p_{\eta})$,
and introduce the Nos\'e Hamiltonian
\begin{eqnarray}
{\cal H}_{\rm N}&=&\sum_i^N\frac{{\bf p}_i^2}{2m_i}+\Phi({\bf r})
+\frac{p_{\eta}^2}{2m_{\eta}}+gk_BT{\eta}
\\
&=& {\cal H}_{\rm T} +gk_BT\eta\;. \label{eq:noseham}
\end{eqnarray}
If one defines the antisymmetric matrix
\begin{equation}
\mbox{\boldmath$\cal B$}=\left[ \begin{array}{cccc}
{\bf 0} & {\bf 0} & {\bf 1}-\mbox{\boldmath$\Delta$} & {\bf 0} \\
{\bf 0} & {\bf 0} &  {\bf 0}& {\bf 1} \\
-{\bf 1}+\mbox{\boldmath$\Delta$} & {\bf 0} & \mbox{\boldmath$\Lambda$} & -{\bf p} \\
 {\bf 0} & -{\bf 1} & {\bf p} & {\bf 0} \end{array}\right]\;,
 \label{eq:nosec}
\end{equation}
and uses this matrix in place of $\mbox{\boldmath ${\cal B}$}^D$
either in~(\ref{eq:b1}) or in~(\ref{eq:genbra}), then,
through the Nos\'e Hamiltonian in Eq.~(\ref{eq:noseham}),
one obtains the desired equations of motion,
\begin{eqnarray}
\dot{\bf r}_i&=&\frac{\bf p}{m_i}\label{eq:nosec1} \\
\dot{\bf p}_i&=&{\bf F}_i+\nabla\mbox{\boldmath$\sigma$}
\cdot\mbox{\boldmath$\lambda$}({\bf r},{\bf p})\label{eq:nosec2}
-{\bf p}_i\frac{p_{\eta}}{m_{\eta}}
\\
\dot{\eta}&=&\frac{p_{\eta}}{m_{\eta}}\\
\dot{p}_{\eta}&=&F_{\eta} \;, \label{eq:nosec4}
\end{eqnarray}
with $F_{\eta}=\sum_i{\bf p}_i^2/m_i-gk_BT$. 

The phase space flow defined {\sl via} {\boldmath$\cal B$} in
Eq.~(\ref{eq:nosec}) conserves the Hamiltonian and any function of
the constraints. The equations of motion
(Eqs.~(\ref{eq:nosec1})-(\ref{eq:nosec4})) have a compressibility
\begin{equation}
\kappa=\sum_{i j}^{2N}\frac{\partial{\cal B}_{ij}}{\partial x_i}
\frac{\partial{\cal H}}{\partial x_j}
=\kappa_c+\kappa_N
\end{equation}
where $\kappa_c$ is given in Eq.~(\ref{eq:k_c}).
The Nos\'e compressibility is $\kappa_N=3N\dot{\eta}$ so that the
total compressibility of Eqs~(\ref{eq:nosec1})-(\ref{eq:nosec4})
is
\begin{equation}
\kappa= -\frac{d\ln|\mbox{\boldmath$Z$}|}{dt}+\beta\frac{d{\cal
H}_T}{dt}\;. \end{equation}
The primitive function of the
compressibility is $w(x)=-\ln|\mbox{\boldmath$Z$}|+\beta{\cal
H}_T$ so that the distribution function 
in the extended phase space is
\begin{eqnarray}
\rho_{\rm ND}({\bf r},{\bf p},\eta,p_{\eta})&=&\delta({\cal H}_{\rm N})
\delta(\mbox{\boldmath$\sigma$})\delta(\dot{\mbox{\boldmath$\sigma$}})
|\mbox{\boldmath$Z$}|e^{-\beta{\cal H}_T}\;.
\label{eq:noseconmeasure}
\end{eqnarray}
One can easily prove that Eq.~(\ref{eq:noseconmeasure}) provides
the distribution of a canonical ensemble with constraints.
Integrating on $\eta$ one has
\begin{equation}
\int d\eta\delta({\cal H}_{\rm N}(\eta))=\int d\eta \delta(\eta)
\left[\frac{d{\cal H}_{\rm N}}{d\eta}\right]^{-1}=\frac{\beta}{g} \;.
\end{equation}
The constant can be absorbed in the normalization and 
the Gaussian integration on $p_{\eta}$ 
can be easily performed so that one obtains
\begin{eqnarray}
\rho_c({\bf r},{\bf p})&= & \delta(\mbox{\boldmath$\chi$})
|\mbox{\boldmath$Z$}|e^{-\beta{\cal H}_0}\;,
\label{eq:rhoc}
\end{eqnarray}
where ${\cal H}_0=\sum_i{\bf p}^2_i/2m_i+\Phi({\bf r})$ is the
Hamiltonian of the physical degrees of freedom.
As an example,
Nos\'e-Dirac flow
will be applied in Appendix~\ref{sec:app} to the sampling of rare events,
and it will be shown how to re-derive
the unbiasing factor first introduced in Ref.~\cite{bm}.

\section{General non-Hamiltonian Dynamics and constraints}
\label{sec:3}

In the case of general non-Hamiltonian dynamics
one cannot derive the generator of time translation
from generalized brackets.
Instead one is led to consider a Liouville
operator~\cite{sllod} of the form
\begin{equation}
iL_{\rm p}=\sum_{i=1}^N
{\bf C}(\{{\bf r},{\bf p}\})\cdot\frac{\partial}{\partial{\bf r}_i}
+{\bf D}(\{{\bf r},{\bf p}\})\cdot\frac{\partial}{\partial{\bf p}_i}
\label{eq:l'}
\end{equation}
defining the time evolution of any arbitrary phase space variable
$a(\{{\bf r},{\bf p}\})$ through $\dot{a}=iL_{\rm p}a$.
The phase space incompressibility condition is usually adopted~\cite{sllod}
\begin{equation}
\frac{\partial{\bf C}}{\partial{\bf r}_i}
+
\frac{\partial{\bf D}}{\partial{\bf p}_i}
=0
\end{equation}
and for simplicity the same will be done here.
In molecular dynamics applications,
Liouville operators, having the same form as that in Eq.~(\ref{eq:l'}),
are used to introduce time dependent perturbations by means of
operators of the form
\begin{equation}
iL_I(t)={\cal F}(t)iL_{\rm p}\;.
\end{equation}
The unperturbed systems is usually subjected
to the action of an operator $iL_0$ which is
instead derivable from some (generalized
or Poisson) bracket with the Hamiltonian.
Accordingly, the total dynamics
is defined {\sl via} the operator 
$iL(t)=iL_0+iL_I(t)$.
In the presence of holonomic constraints,
for example describing rigid  molecules, 
the formalism of Ref.~\cite{sergi-dirac},
for the Hamiltonian case,
or of the previous section, for non-Hamiltonian dynamics,
can be used to define an operator
$iL_0^D$ having the constraints as conserved 
quantities.
The problem is that $iL_{\rm p}$ and $iL_I(t)$ as such
do not preserve the constraints and could lead
to spurious term in the linear response derivation,
as it will be shown in the following.
This feature of the formalism is not desiderable
since,
in actual molecular dynamics calculations, the algorithms
enforcing the constraints is used in the presence of the perturbation
determined by $iL_I(t)$~\cite{allen,edberg,marechal,travis}
so that this perturbation does not violate the constraints in practice.
The conclusion is that, in order to set up
a correct formalism, one must project out
the dynamics that $iL_{\rm p}$ and $iL_I(t)$ spuriously
impose on the constraints.
To this aim, 
by using a simple extension
of Dirac's recipe to general 
non-Hamiltonian Liouville operators,
one can define 
$iL_I^D(t)$ as follows
\begin{eqnarray}
iL_I^D(t)a&=&iL_I(t)a\nonumber \\
&-&\sum_{\alpha\beta}\{a,\chi_{\alpha}\}
({\bf C}^{-1})_{\alpha\beta}(iL_I(t)\chi_{\beta})\nonumber\\
&=&{\cal F}(t)\left[
iL_{\rm p}a\right.\nonumber \\
&-&\left.\sum_{\alpha\beta}\{a,\chi_{\alpha}\}
({\bf C}^{-1})_{\alpha\beta}(iL_{\rm p}\chi_{\beta})\right]\nonumber \\
&=&{\cal F}(t)iL_{\rm p}^Da
\;,
\label{eq:gendirac}
\end{eqnarray}
where  $a$ is an arbitrary phase space variable,
$\{a,\chi_{\alpha}\}$ is the Poisson bracket
and ${\bf C}$ is derived by means of Eqs.~(\ref{eq:c})
and~(\ref{eq:c-1}).
As in the Hamiltonian case~\cite{sergi-dirac}
or in that of the non-Hamiltonian bracket, $iL_I^D(t)$ 
preserves any function of the constraints, by construction.
Equation~(\ref{eq:gendirac}) generalizes Dirac's theory to arbitrary
non-Hamiltonian phase space flows.

By means of $iL_I^D(t)$ one can set up 
the correct formalism for the linear response
of systems with holonomic constraints subject to
a non-Hamiltonian time-dependent perturbation.
For simplicity, the case in which the unperturbed dynamics
of the constrained system is Hamiltonian will be considered
in the following.
In this situation, the unperturbed system has a conserved Hamiltonian
${\cal H}_0$, a Liouville operator
$iL_0^D=\sum_i{\cal B}_{ij}^D\partial/\partial x_i$,
with $\mbox{\boldmath$\cal B$}^D$ defined by Eq.~(\ref{B^D}),
and an equilibrium distribution function given
by $\rho_e=||\mbox{\boldmath$Z$}||\delta(\mbox{\boldmath$\chi$})\delta({\cal H}_0)$.
The Liouville equation in the presence of the perturbation
is
\begin{equation}
\frac{\partial\rho}{\partial t}=-i{\cal L}_0^D\rho-iL_I^D(t)\rho\;,
\end{equation}
where~\cite{sergi-dirac} $i{\cal L}_0^D=iL_0^D+\kappa_c$.
One can consider $\rho=\rho_e+\delta\rho$ and to linear order
\begin{equation}
\delta\rho(t)=-\int_0^td\tau{\cal F}(\tau)e^{-i{\cal L}_0^D(t-\tau)}iL_{\rm p}^D\rho_e
\;.
\end{equation}
The nonequilibrium average of 
$\delta b(t)=b(t)-\langle b\rangle_{eq}$
for any phase space variable is then
\begin{equation}
\overline{\delta b}(t)=\int_0^td\tau{\cal F}(\tau)\phi(t-\tau)\;,
\end{equation}
where
\begin{equation}
\phi(t)=-\int dx b(t)iL_{\rm p}^D\rho_e
\end{equation}
is the response function and $b(t)=\exp[iL_0^Dt]b(0)$ since the compressibility
$\kappa_c$ disappears when integrating by parts 
$\exp[i{\cal L}_0^Dt]$~\cite{sergi-dirac,b2}.
Now, in evaluating the action of $iL_{\rm p}^D$ on $\rho_e$
one can take full advantage of the fact that
$iL_{\rm p}^D\delta(\mbox{\boldmath$\chi$})=0$.
Had one used $iL_{\rm p}$ instead, spurious terms
would have appeared. Thus
\begin{eqnarray}
iL_{\rm p}^D\rho_e=\rho_e
\left[iL_{\rm p}^D\ln||\mbox{\boldmath$Z$}||
-\beta (iL_{\rm p}^D{\cal H}_0)\right]\;,
\end{eqnarray}
where it has been used the fact that
$iL_{\rm p}^D\delta({\cal H}_0)\approx-\beta\delta({\cal H}_0)(iL_{\rm p}^D{\cal H}_0)$
in the thermodynamic limit~\cite{bishop}.
Hence, the response function for constrained systems takes the form
\begin{equation}
\phi(t)=\beta\langle\left(iL_{\rm p}^D{\cal H}_0
-\beta^{-1}iL_{\rm p}^D\ln||\mbox{\boldmath$Z$}||\right)\rangle_{eq} \;.
\end{equation}
The correction factor arising from $iL_{\rm p}^D\ln||\mbox{\boldmath$Z$}||$
disappears if 
\begin{equation}
iL_{\rm p}=\sum_i{\bf D}_i\partial/\partial{\bf p}_i
\label{eq:approxilp}
\end{equation}
but for general equations of 
motion~\cite{ladd,sllod,evansmorriss,neqtuck,allen,edberg,marechal,travis}
it must be considered.

\section{Conclusions}

The extension of Dirac's formalism allows one
to treat correctly systems with holonomic constraints
undergoing non-Hamiltonian dynamics.
Non-Hamiltonian dynamics can be derived from generalized brackets
or it can be more general and not be derivable from any brackets:
both cases have been treated.
Using generalized brackets, a Nos\'e-Dirac
phase space flow has been introduced
and applied to derive the unbiasing factor
when constraints are used to sample rare
events. It has been shown how to generalize Dirac's
recipe when the dynamics cannot be obtained from 
brackets. Linear response theory of system with
holonomic constraints subjected to general non-Hamiltonian perturbation
has been illustrated.
The use of Dirac's formalism makes spurious terms
disappear from the response function. However, a correction coming from
the measure of constrained phase space is present 
in general cases.  Further work is required in order to assess
the importance of this correction in numerical
calculations on condensed matter systems.

Equilibrium statistical mechanics
and linear response theory  of systems with nonholonomic constraints 
remain to be addressed.
However, as a consequence of the analysis presented in this
paper, it can be suggested that
a formalism, suitable for the linear
response of such systems, must project
the spurious time evolution of the constrained variables
out of both unperturbed and perturbed dynamics. 

\vspace{1cm}
\begin{flushleft}
{\bf Acknowledgments}
\end{flushleft}
The author is grateful to Professor Raymond
Kapral for many discussions.

\appendix

\section{Nos\'e-Dirac flow and rare events}\label{sec:app}

Often one is interested in the calculation of conditional averages
of some phase space function $a({\bf r})$
\begin{equation}
\langle a({\rm r})\rangle_{\rm cond}=\frac{\langle a({\rm r})
\delta(\mbox{\boldmath$\xi$}({\bf r})-\mbox{\boldmath$\xi$}^{\ddag})\rangle_{NVT}}
{\langle \delta(\mbox{\boldmath$\xi$}({\bf r})-\mbox{\boldmath$\xi$}^{\ddag})\rangle_{NVT}}
\label{eq:condav}
\end{equation}
where $\langle\ldots\rangle_{NVT}$ stands for an equilibrium average
in the canonical ensemble.
Molecular dynamics can be used to perform calculation
where the condition 
$\mbox{\boldmath$\sigma$}({\bf r})=\mbox{\boldmath$\xi$}-\mbox{\boldmath$\xi$}^{\ddag}=0$
is treated as a holonomic constraint. However this
automatically brings in a constraints on the time variation
$\dot{\mbox{\boldmath$\sigma$}}({\bf r},\dot{\bf r})$
so that using the Nos\'e-Dirac flow introduced before
one would get a constrained average,
defined by
\begin{eqnarray}
\langle a({\bf r}) \rangle _{\xi^{\ddag}}&=& \frac{\langle a({\bf r})
||\mbox{\boldmath$Z$}||\delta(\mbox{\boldmath $\chi$})
\rangle_{NVT}}
{\langle ||\mbox{\boldmath$Z$}||\delta(\mbox{\boldmath $\chi$})
\rangle_{NVT}}
\label{eq:constrav}
\end{eqnarray}
where
$\delta(\mbox{\boldmath$\chi$})=\delta(\mbox{\boldmath$\sigma$})
\delta(\dot{\mbox{\boldmath$\sigma$}})$.

The relation between conditional average~(\ref{eq:condav}) 
and constrained averages~(\ref{eq:constrav}) has been originally
given in Ref.~\cite{bm}. In the present context, it is simply 
remarked that, since the formal manipulations are performed
in the canonical ensemble, in order to be rigorous
one needs the Nos\'e-Dirac flow to have the correct distribution
function given in Eq.~(\ref{eq:rhoc}).
Having said that, one just needs the results of Ref~\cite{andersendemon},
which show that
$\int d^N{\bf p}||\mbox{\boldmath$Z$}||\delta(\dot{\mbox{\boldmath$\sigma$}})
\propto ||\mbox{\boldmath$Z$}||^{1/2}$,
in order to re-write the constrained average as
\begin{eqnarray}
\langle a({\bf r}) \rangle _{\xi^{\ddag}}&=& \frac{\langle a({\bf r})
||\mbox{\boldmath$Z$}||^{1/2}\delta(\mbox{\boldmath $\sigma$})
\rangle_{NVT}}
{\langle ||\mbox{\boldmath$Z$}||^{1/2}\delta(\mbox{\boldmath $\sigma$})
\rangle_{NVT}}\;.
\label{eq:constrav2}
\end{eqnarray}
From this, one immediately obtains the result of Ref.~\cite{bm} 
\begin{eqnarray}
\langle a({\bf r})\rangle_{cond}
&=&
\frac{\langle |\mbox{\boldmath $Z$}|^{-1/2} a({\bf r})
\rangle_{\xi^{\ddagger}}}
{\langle |\mbox{\boldmath $Z$}|^{-1/2}\rangle_{\xi^{\ddag}}}\;.
\label{unbiasing-xi}
\end{eqnarray}


\end{document}